\documentclass[12pt]{article}

\usepackage{amssymb}
\usepackage{amsmath}
\usepackage{amscd}
\usepackage{latexsym}

\usepackage{graphicx}
\usepackage{cite}

\newcommand{\be}{\begin{equation}}
\newcommand{\ee}{\end{equation}}
\newcommand{\Dlt}{\Delta}

\newcommand{\prt}{\partial}
\newcommand{\br}{{\bf r}}

\newcommand{\bt}{\beta}
\newcommand{\vp}{\varphi}
\newcommand{\ep}{\varepsilon}
\newcommand{\al}{\alpha}
\newcommand{\ra}{\rightarrow}
\newcommand{\sgm}{\sigma}

\newcommand{\gm}{\gamma}
\newcommand{\om}{\omega}

\newcommand{\rgl}{\rangle}
\newcommand{\lgl}{\langle}

\begin{document}

\begin{center}

{\Large{\bf Critical indices from self-similar root approximants}  \\ [5mm]

S. Gluzman and V.I. Yukalov$^a$}  \\ [3mm]

{\it
Bogolubov Laboratory of Theoretical Physics, \\
Joint Institute for Nuclear Research, Dubna 141980, Russia \\ [3mm]
}

\end{center}

\vskip 2cm

\begin{abstract}
The method of self-similar root approximants has earlier been shown to provide 
accurate {\it interpolating} formulas for functions for which small-variable expansions 
are given and the behaviour of the functions at large variables is known. Now this
method is generalized for the purpose of {\it extrapolating} small-variable expansions 
to the region of finite and large variables, where the sought function exhibits critical 
behaviour. The procedure of calculating critical indices is formulated and illustrated by
a variety of physical problems.      

\end{abstract}

\vskip 3cm

$^a${\bf Corresponding author}: V.I. Yukalov

\vskip 1mm
{\bf E-mail}: yukalov@theor.jinr.ru

\newpage

\section{Formulation of problem}

A real function $f(x)$ of a real variable $x$ is said to exhibit critical behaviour, with a
critical index $\beta$, at a finite critical point $x_c$, when in the vicinity of this point
it behaves as
\be
\label{1}
 f(x) \simeq B (x_c - x)^\bt \qquad ( x \ra x_c - 0 ) \;  .
\ee
The function can tend to infinity, if the critical index $\beta$ is negative, or to zero,
if this index is positive. 

The critical behaviour can also occur at infinity, where the function behaves as
\be
\label{2}
  f(x) \simeq B x^\bt \qquad ( x \ra \infty) \;  ,
\ee
with the critical index $\beta$. Respectively, the function can tend to infinity, if $\beta$ 
is positive and to zero, if $\beta$ is negative. The critical behaviour at infinity can formally 
be interpreted as the case, where the critical point is located at infinity. 

Critical phenomena are widespread in physics. And it is an important problem of defining
the related critical indices. However, for realistic physical systems it is often impossible 
to get exact solutions, but the sole thing one can do is to resort to perturbation theory for
obtaining the behavior of the sought function at small variable,
\be
\label{3}
 f(x) \simeq f_k(x) \qquad ( x \ra 0) \;  ,
\ee
where the function is approximated by an expansion
\be
\label{4}
 f_k(x) = f_0(x) \left ( 1 + \sum_{n=1}^k a_n x^n \right ) \;  ,
\ee
with $f_0(x)$ being a known factor. Without loss of generality, we may assume that
this prefactor has the form
\be
\label{4a}
 f_0(x) = A x^\al \;  .
\ee
Such expansions are usually asymptotic and strongly divergent not allowing for their use 
at finite values of the variable.  

In order to understand whether the function possesses critical behaviour, it is, first of all,
necessary to extrapolate the asymptotic expansion (\ref{4}) to finite and even large values
of the variable. Such an extrapolation can be accomplished by means of Pad\'{e} and Borel
summation or other techniques usually requiring the knowledge of many expansion terms
\cite{Baker_1,Kleinert_2}. In addition, these summation techniques not always are applicable, 
as is discussed in Refs. \cite{Gluzman_3,Gluzman_4}. How would it be possible to obtain 
reliable results for the critical indices employing a small number of terms in the asymptotic 
expansion? 

An efficient method of summation of divergent series has been developed in the frame of
self-similar approximation theory \cite{Yukalov_5,Yukalov_6,Yukalov_7,Yukalov_8}. A 
variant of this method, called self-similar root approximants, has been shown by a number 
of problems \cite{Gluzman_9,Yukalov_10,Yukalov_11,Gluzman_12,Gluzman_13,Yukalov_14}
to provide accurate {\it interpolation} formulae, when both the small-variable expansion and 
the large-variable behaviour are known. The accuracy of such an interpolation was 
demonstrated to be not worse and in many cases better than that of two-point Pad\'{e} 
approximants, when these could be defined.  

In the present paper, we generalize the method of self-similar root approximants for the 
{\it extrapolation} purpose. We consider the situation, when only small-variable expansions
are given, and one needs to extrapolate such expansions to finite, or even infinite, values
of the variable, not knowing the large-variable behaviour of the sought functions. The main 
attention will be payed to the problem of calculating the critical indices. 

According to the general theory \cite{Yukalov_11,Gluzman_12,Gluzman_13,Yukalov_14},
a self-similar root approximant, based on the small-variable expansion (\ref{4}), has the 
form
\be
\label{5}
 \frac{f_k^*(x,n_k)}{f_0(x)} = \left ( \left ( 
( 1 + A_1 x)^{n_1} + A_2 x^2 \right )^{n_2} + \ldots + A_k x^k \right )^{n_k} \; ,
\ee
in which all parameters $A_j$ are found from the comparison of the like orders of the 
re-expansion of equation (\ref{5}) in powers of $x$ with the given expansion (\ref{4}).
This method of equating the like powers of $x$ is sometimes called the 
accuracy-through-order procedure. The internal powers are defined as
\be
\label{6}
 n_j = \frac{j+1}{j} \qquad ( j = 1,2,\ldots,k-1) \;  ,
\ee
while the external power $n_k$ plays the role of a control function to be determined from 
additional conditions.      

If the large-variable power $\beta$ in equation (\ref{2}) were known, then we could compare
the latter equation with the large-variable behaviour of the root approximant (\ref{5}) being
\be
\label{7}
 f_k^*(x,n_k) \simeq B_k x^{\al+ k n_k} \;  ,
\ee
where $\alpha$ is introduced in equation (\ref{4a}) and
\be
\label{8}
 B_k = A \left ( \left ( \left ( A_1^{n_1} + A_2 \right )^{n_2} + 
A_3 \right )^{n_3} + \ldots + A_k \right )^{n_k} \;  .
\ee
This comparison yields the relation
\be
\label{9}
   \al + k n_k = \bt
\ee
defining the external power 
\be
\label{10}
 n_k = \frac{\bt-\al}{k} \;  ,
\ee
provided $\beta$ is known. This way of defining the external power has been used
in our previous papers on the application of the root approximants as interpolating
formulae. When several terms in the large-variable behaviour are known, then the 
related powers prescribe the values for the corresponding number of external 
powers $n_j$.

But now we consider the situation where the large-variable behaviour of the function
is not known, hence $\beta$ is not given. Moreover, the critical behaviour can happen
at a finite value $x_c$ of the variable $x$. The development of the method for defining 
the critical index $\beta$ by employing self-similar root approximants is the aim of the 
present paper.

\section{Method of defining critical indices}

Suppose we can construct several root approximants $f_k^*(x,n_k)$, in which the 
external power $n_k$ plays the role of a control function. Following the idea of
self-similar approximation theory \cite{Yukalov_5,Yukalov_6,Yukalov_7}, it is possible
to treat the sequence $\{f_k^*(x,n_k)\}$ as a trajectory of a dynamical system, with 
the approximation order $k$ playing the role of discrete time. A discrete-time 
dynamical system is called cascade. Here it is termed the approximation cascade, 
since its trajectory consists of the sequence of approximants. The cascade velocity 
is defined by the Euler discretization formula
\be
\label{11}
 V_k(x,n_k) = f_{k+1}^*(x,n_k) - f_k^*(x,n_k) +
(n_{k+1}- n_k ) \; \frac{\prt}{\prt n_k} \; f_k^*(x,n_k) \;  .
\ee
The effective limit of the sequence $\{f_k^*(x,n_k)\}$ corresponds to the fixed point
of the cascade, where the cascade velocity tends to zero, as $k$ tends to infinity. 
Having a finite number of approximants, the cascade velocity is not necessarily 
tending to zero, but certainly has to diminish for the sequence being convergent. Thus,
the control functions $n_k = n_k(x)$, controlling convergence, are defined as the 
minimizers of the absolute value of the cascade velocity 
\cite{Yukalov_5,Yukalov_6,Yukalov_7,Yukalov_8}:
\be
\label{12}
  | V_k(x,n_k(x)) | = \min_{n_k}  | V_k(x,n_k) | \; .
\ee
      
A finite critical point $x_k^c$, in the $k$-th approximation, exists if the equation
\be
\label{13}
[ f_k^*(x_k^c,n_k) ]^{1/n_k} = 0 \qquad ( 0 < x_k^c < \infty )
\ee
enjoys a finite solution for $x_k^c = x_k^c(n_k)$. Then the critical index in the $k$-th
approximation is
\be
\label{14}
 \bt_k = \lim_{x\ra x_k^c} n_k(x) \qquad  ( 0 < x_k^c < \infty ) \;  .
\ee

When we are studying the critical behaviour at infinity, which we denote as $x_c \sim \infty$,
keeping in mind that this case formally corresponds to the critical point at infinity, then
the critical index is
\be
\label{15}
 \bt_k = \al + k \lim_{x\ra \infty} n_k(x) \qquad  ( x_c \sim \infty ) \; ,
\ee
where $\alpha$ is defined in equation (\ref{4a}).

Thus the critical indices are defined, provided the control functions $n_k(x)$ are found.
However, the minimization of the cascade velocity (\ref{11}) poses some problems. First
of all, equation (\ref{12}) contains two control functions, $n_{k+1}$ and $n_k$. Hence it
is impossible to find two solutions from one equation. But it is possible to simplify the 
problem, when minimizing velocity (\ref{11}), so that to get one equation for one control 
function. This is admissible to accomplish in two ways. 

For instance, keeping in mind that $n_{k+1}$ is close to $n_k$, equation (\ref{12}) reduces 
to the {\it minimal difference condition}
\be
\label{16}
  \min_{n_k} \left | f_{k+1}^*(x,n_k) - f_k^*(x,n_k) \right | \qquad
(k=1,2,\ldots ) \; .
\ee
In particular, one can look for a solution $n_k = n_k(x)$ of the equation
\be
\label{17}
  f_{k+1}^*(x,n_k) - f_k^*(x,n_k) = 0 \; .
\ee
If the latter does not possess a solution for $n_k$, one has to return to form (\ref{16}). 

In general, when nothing is known on the form of the sought function $f(x)$, the control 
functions $n_k$, being the solutions of equation (\ref{17}), depend on the variable $x$.
But when we are looking for a function $f(x)$ in the vicinity of its critical point $x_c$, 
where the function $f(x)$ acquires form (\ref{1}), the control functions are to be treated 
as the limits of $n_k(x)$ for $x \ra x_c$. So that the control functions $n_k$, characterizing 
the critical behaviour of $f(x)$ near a critical point $x_c$, become the numbers $n_k(x_c)$. 
In what follows, writing for short $n_k$, we assume $n_k = n_k(x_c)$.    

In the vicinity of a finite critical point, the function $f_k^*$ behaves as
\be
\label{18}
 f_k^*(x,n_k) \simeq f_0(x) \left ( 1 - \; \frac{x}{x_k^c} \right )^{n_k} \qquad
( x \ra x_k^c - 0 ) \;  .
\ee
Then condition (\ref{17}) becomes
\be
\label{19}
  x_{k+1}^c(n_k) - x_k^c(n_k)= 0 \qquad (0 < x_k^c < \infty ) \; .
\ee

When the critical behaviour occurs at infinity, then it is convenient to introduce the
control function
\be
\label{20}
s_k \equiv k n_k \;  ,
\ee
so that the large-variable behaviour of the root approximants reads as
\be
\label{21}
 f_k^*(x,s_k) \simeq B_k(s_k) x^{\al + s_k} \qquad ( x \ra \infty ) \;  .
\ee
As a result, the minimal difference condition
\be
\label{22}
 f_{k+1}^*(x,s_k) -  f_k^*(x,s_k) = 0
\ee
leads to the equation
\be
\label{23}
 B_{k+1}(s_k) - B_k(s_k) = 0 \qquad ( x_k^c \sim \infty ) \;  .
\ee

The other equation for defining control functions follows from the minimal velocity 
condition (\ref{12}) by keeping in mind that $f_{k+1}^*$ is close to $f_k^*$ and usually 
$n_{k+1}$ does not exactly coincide with $n_k$, because of which one has to consider 
the {\it minimal derivative condition}  
\be
\label{24}
 \min_k \left | \frac{\prt}{\prt n_k} \; f_k^*(x,n_k) \right | \qquad
(k=1,2,\ldots ) \;  .
\ee
In particular, one can look for the solution of the equation
\be
\label{25}
 \frac{\prt}{\prt n_k} \; f_k^*(x,n_k) = 0 \;  .
\ee

However, the minimal derivative condition cannot be applied directly, when the sought
function exhibits critical behaviour, where the function either diverges or tends to zero. 
To apply this condition, it is necessary to extract from the function nondivergent  parts. 
For example, if the critical point is finite, one can study the residue of the function 
$\partial \ln f_k^*/ \partial n_k$, for which we have
$$
 \lim_{x\ra x_k^c} ( x_k^c - x) \;  \frac{\prt}{\prt n_k} \; \ln f_k^*(x,n_k) 
= n_k \; \frac{\prt x_k^c}{\prt n_k} \;  .
$$
Therefore, instead of equation (\ref{25}), we get the condition
\be
\label{26}
 \frac{\prt x_k^c}{\prt n_k} = 0  \qquad (0 < x_k^c < \infty ) \; .
\ee
And when the critical behaviour occurs at infinity, then we can consider
the limit
$$
 \lim_{x\ra\infty} \; \frac{f_k^*(x,s_k)}{x^{\al+s_k} } = B_k(s_k) \;  ,
$$
for which equation (\ref{25}), defining control functions, reduces to the form
\be
\label{27}
 \frac{\prt B_k(s_k)}{\prt s_k} = 0 \qquad (x_k^c \sim \infty) \; .
\ee
  
To better explain the suggested techniques, let us consider a simple example, when
the sought function leads to the small-variable expansion
\be
\label{28}
 f(x) \simeq 1 + a_1 x + a_2 x^2 \qquad ( x\ra 0 ) \;  .
\ee
The first-order root approximant is
\be
\label{29}
 f_1^*(x,n_1) = (1 + A x)^{n_1} \;  ,
\ee
with 
$$
A = A(n_1) = \frac{a_1}{n_1}
$$
found from the accuracy-through-order procedure. Note that this form (\ref{29}) reminds
us the Sommerfeld formula used in nuclear physics \cite{Sommerfeld_15}. Expression 
(\ref{29}) possesses a finite critical point, provided there exists a finite positive value
\be
\label{A1}
 x_1^c(n_1) = -\; \frac{1}{A(n_1)} = -\; \frac{n_1}{a_1} \;  .
\ee

The second-order root approximant reads as
\be
\label{30}
 f_2^*(x,n_2) = \left (  (1 + A_1 x)^2 + A_2 x^2 \right )^{n_2} \; ,
\ee
with the parameters
$$
 A_1 = A_1(n_2) = \frac{a_1}{2n_2} \; , \qquad
 A_2 = A_2(n_2) = \frac{a_1^2(1-2n_2)+4a_2n_2}{4n_2^2} \;  .
$$
Now the finite critical point is given by a positive value
\be
\label{A2}
 x_2^c(n_2) = \frac{-A_1(n_2) \pm\sqrt{-A_1(n_2)} }{A_1^2(n_2) + A_2(n_2)} \;  .
\ee
The minimal difference condition (\ref{17}), in the form
\be
\label{A3}
 f_2^*(x,n_1) - f_1^*(x,n_1) = 0 \;  ,
\ee
is equivalent to the condition
\be
\label{A4}
 x_2^c(n_1) - x_1^c(n_1) = 0 \;  ,
\ee
which yields
\be
\label{A5}
 n_1 = \frac{a_1^2}{a_1^2-2a_2} \;  .
\ee
Hence the first-order critical point is
\be
\label{A6}
 x_1^c(n_1) = -\; \frac{n_1}{a_1} = \frac{a_1}{2a_2-a_1^2} \;  .
\ee
In what follows, keeping in mind the minimal difference condition, we shall write, 
for simplicity, $x_1^c(n_1) = x_2^c(n_1) \equiv x_c$.

The corresponding root approximant (\ref{29}) acquires the form
\be
\label{A7}
 f_1^*(x) = \left ( 1 - \; \frac{x}{x_c}\right )^{\bt_1} \qquad (0\leq x \leq x_c) \;  ,
\ee
with the first-order critical index
\be
\label{A8}
 \bt_1 = n_1 = \frac{a_1^2}{a_1^2-2a_2} \;  .
\ee

The second-order critical index $\beta_2 = n_2$ is defined by the condition
\be
\label{A9}
 \frac{\prt}{\prt n_2} \; x_2^c(n_2) = 0 \;  .
\ee

When there is no finite critical point, but the critical behaviour happens at infinity, 
then we have to consider the large-variable asymptotic forms for the root 
approximants, with the substitution (\ref{20}). The first-order approximant gives 
\be
\label{A10}
 f_1^*(x,s_1) \simeq B_1(s_1) x^{s_1} \qquad (x\ra\infty) \;  ,
\ee
where
$$
  B_1(s_1)  = A(s_1)^{s_1} \;  .
$$
While the second-order approximant leads to
\be
\label{A11}
 f_2^*(x,s_2) \simeq B_2(s_2) x^{s_2} \qquad (x\ra\infty) \;  ,
\ee
with
$$
 B_2(s_2) = \left [ A_1(s_2)^2 + A_2(s_2) \right ]^{s_2/2} \;  .
$$
The first-order critical index $\beta_1 = s_1$ is found form the condition
\be
\label{A12}
  B_2(s_1) - B_1(s_1)  =  0 \qquad (\bt_1 = s_1) \;  .
\ee
And the second-order critical index $\beta_2 = s_2$ can be obtained form the 
condition
\be
\label{A13}
 \frac{\prt B_2(s_2)}{\prt s_2}  = 0  \qquad (\bt_2 = s_2)\;   .
\ee
The final answer can be presented as the average
\be
\label{A14}
 \bt^* = \frac{1}{2} \; (\bt_1 + \bt_2) \pm \frac{1}{2} \; | \bt_1 -\bt_2 | \;  .
\ee

In the similar way, one can proceed to higher orders. However, our aim here is 
to show that even in the lower orders we get rather accurate critical indices, which
is demonstrated for various problems in Secs. 4 to 12. The existence of numerical
convergence for higher orders is illustrated in Sec. 14. Section 15 concludes.

\section{Comparison with method of Pad\'{e} approximants}

Asymptotic series are often approximated by Pad\'{e} approximants. The latter is denoted
as $P_{M/N}(x)$ implying the ratio of a polynomial of order $M$ with respect to $x$ to 
a polynomial of order $N$. The coefficients of the polynomials are defined by comparing
the like orders of expansions in powers of $x$ of the Pad\'{e} approximant and of the 
sought function $f_k(x)$. To emphasize this fact, one often denotes the related  Pad\'{e} 
approximant as $P_{M/N}(x,f)$. These approximants provide the best approximation in the 
class of rational functions \cite{Baker_1}. However the functions in the vicinity of their 
critical points, in general, are non-rational. Therefore the direct use of Pad\'{e} approximants 
for functions exhibiting critical behaviour is impossible. Really, a Pad\'{e} approximant 
$P_{M/N}$ can have a pole that could be associated with a finite critical point, but the 
related critical index would be an integer, while usually critical indices are not integers. 
The same concerns the large-variable behaviour of $P_{M/N}(x)$, where the power of 
$x$ is always an integer $M - N$. 

The problem can be circumvented by means of the Dlog-Pad\'{e} method 
\cite{Baker_1,He_15}. Then, instead of the function $f(x)$, one considers the function
\be
\label{B1} 
 g(x) \equiv \frac{d}{dx} \; \ln f(x) \;  .
\ee
Near a finite critical point $x_c$, where the function $f(x)$ is of the form (\ref{1}),
function (\ref{B1}) behaves as
\be
\label{B2}
 g(x) \simeq \frac{\bt}{x-x_c} \qquad ( x\ra x_c -0 ) \;  .
\ee
Hence the pole here defines the critical point $x_c$, while the critical index is given by
the residue
\be
\label{B3}
 \bt = \lim_{x\ra x_c} (x-x_c) g(x) \;  .
\ee
  
When the sought function is represented by expansion (\ref{4}), function (\ref{B1}) takes
the form
\be
\label{B4}
\frac{d}{dx} \; \ln f_k(x)  = \frac{\al}{x} + g_k(x) \;  ,
\ee
in which the second term can be expanded in powers of $x$ yielding 
\be
\label{B5}
  g_k(x) \simeq \sum_{n=0}^k a_n' x^n \qquad ( x \ra 0 ) \; ,
\ee
with the coefficients $a'_n$ expressed through $a_n$. The finite series (\ref{B5}) can be 
extrapolated by Pad\'{e} approximants $P_{M/N}(x,g)$, with $M + N \leq k$. The pole 
nearest to zero of the latter approximant defines the critical point $x_c$, and the critical 
index is given by the residue
\be
\label{B6}
 \bt_{M/N} = \lim_{x\ra x_c} (x-x_c) P_{M/N}(x,g) \;  .
\ee

For a function $f(x)$, with the critical behaviour at infinity, where it has form (\ref{2}),
function (\ref{B1}) behaves as
\be
\label{B7}
g(x) \simeq \frac{\bt}{x} \qquad ( x\ra\infty) \;   .
\ee
This implies that the critical index can be obtained from the limit
\be
\label{B8}
 \bt = \lim_{x\ra\infty} x g(x) \;   .
\ee

It is clear that extrapolating the function $g_k(x)$ by Pad\'{e} approximants $P_{M/N}(x,g)$,
not all $M$ and $N$ are permitted, since in the large-variable limit the Pad\'{e} approximants
$P_{M/N}(x,g)$ exhibit different behaviour depending on the relation between $M$ and $N$:
\begin{eqnarray}
\nonumber
\lim_{x\ra\infty}  P_{M/N}(x,g) = \left \{ \begin{array}{ll}
0 , ~ & ~ M < N \\
const  , ~ & ~ M = N \\
\infty , ~ & ~ M > N \; .
\end{array} \right.
\end{eqnarray}
In order that the critical index (\ref{B8}) be finite, it is necessary to take the approximants
$P_{N/N+1}(x,g)$, so that the corresponding approximation 
\be
\label{B9}
 \bt_{N/N+1}(x,g) = \lim_{x\ra\infty} x P_{N/N+1}(x,g)
\ee
be finite. 

As is known \cite{Baker_1}, for a given expansion of order $k$, one can construct the 
whole table of Pad\'{e} approximants. This means that defining the critical indices 
through the Dlog-Pad\'{e} method is not a uniquely defined procedure. And different 
Pad\'{e} approximants can result in basically different values. Then it is not clear 
which of these quantities to prefer.  

For illustration, let us again consider the case of expansion (\ref{28}). We define 
function (\ref{B5}) and construct the related Pad\'{e} approximants $P_{M/N}(x,g)$, with 
$M+N\leq 2$. The simplest Pad\'{e} approximant here is
\be
\label{B10}
 P_{0/1}(x,g) = \frac{b_0}{1+ b_1 x} \;  ,
\ee
with the parameters
$$
b_0 = a_1 \; , \qquad b_1 = \frac{a_1^2-2a_2}{a_1} \;   .
$$
The pole of approximant (\ref{B10}) yields the critical point
\be
\label{B11}
 x_c = - \; \frac{1}{b_1} = \frac{a_1}{2a_2-a_1^2} \;  .
\ee
The critical index
$$
\bt_{0/1} = \lim_{x\ra x_c} ( x - x_c ) P_{0/1}(x,g)
$$
becomes
\be
\label{B12}
\bt_{0/1} = \frac{b_0}{b_1}  = \frac{a_1^2}{a_1^2-2a_2} \;   .
\ee
In this case, comparing equations (\ref{B11}) with (\ref{A6}) and (\ref{B12}) with 
(\ref{A8}), we notice that the simplest Dlog-Pad\'{e} approximation $P_{0/1}$ gives 
the critical point and critical index coinciding with those of the first root 
approximation. However, in the frame of the Dlog-Pad\'{e} method, the choice of 
a particular Pad\'{e} approximant is not uniquely defined, since there are several 
possibilities of choosing such Pad\'{e} approximants. Thus, we can take 
\be
\label{B13}
 P_{1/1}(x,g) = \frac{c_0 + c_1x}{1+c_2 x} \;  ,
\ee
with the parameters
$$
 c_0 = a_1 \; , \qquad c_1 = \frac{4a_2 - a_1^2}{2a_2 - a_1^2}\; a_2 \; , \qquad
c_2 = \frac{3a_2 - a_1^2}{2a_2 - a_1^2}\; a_1 \;  .
$$
Then the critical point is
\be
\label{B14}
 x_c' = -\; \frac{1}{c_2} = -\; \frac{2a_2 - a_1^2}{a_1(3a_2 - a_1^2)} \;  .
\ee
And the critical index
$$
\bt_{1/1} = \lim_{x\ra x_c'} ( x - x_c') P_{1/1}(x,g)
$$
becomes
\be
\label{B15}
 \bt_{1/1} = \frac{a_1^2(4a_2 - a_1^2)-4a_2^2}{2a_2 - a_1} \;  .
\ee
These values can be quite different from the previously found. The critical points
(\ref{B14}) and (\ref{B11}) are connected by the relation
\be
\label{B16}
 x_c' = -\; \frac{1}{(3a_2-a_1^2)x_c} \;  ,
\ee
while the critical indices (\ref{B15}) and (\ref{B12}), by the relation
\be
\label{B17}
 \bt_{1/1} = ( 2a_2 - a_1^2)^2 \; \frac{\bt_{0/1}}{a_1^2} \;  .
\ee
As is seen from equality (\ref{B16}), under the condition $3 a_2 > a_1^2$, the critical 
points $x_c$ and $x'_c$ are of different signs. Hence, if one of them is positive, the 
other is negative, that is, does not exist as a physically acceptable solution. And the 
critical indices $\bt_{1/1}$ and $\bt_{0/1}$ can be quite different. This is contrary to 
the case of root approximants, where the second-order approximations for the critical 
point (\ref{A2}) and the critical index (\ref{A9}) do not contradict the first-order values.

In the following sections, by treating a number of concrete examples, we show that the 
Dlog-Pad\'{e} approximation, based on approximant (\ref{B13}), either does not possess
physical solutions or is worse than the values given by the root approximants. Moreover,
analyzing the numerical convergence of solutions in Sec. 14, we demonstrate that the 
root approximants yield a sequence of solutions converging to the exact value of a critical 
index, while the sequence, based on the Dlog-Pad\'{e} method, is divergent and results
in unphysical solutions.

\section{Susceptibility of two-dimensional Ising model}

Consider the two-dimensional Ising model characterized by the Hamiltonian
\be
\label{31}
  \hat H = -\; \frac{J}{2} \sum_{\lgl ij\rgl} s_i^z s_j^z \qquad 
\left ( s_j^z \equiv \frac{S_j^z}{S} \right ) 
\ee
on a square lattice, with the ferromagnetic interaction of nearest neighbours, for 
spins $S_j^z = \pm 1/2$. The dimensionless interaction parameter is defined as
\be
\label{32}
 g \equiv \frac{J}{k_B T} \;  .
\ee
The susceptibility is known \cite{Baxter_16} to diverge at a critical point 
\be
\label{33}
g_c = \frac{1}{2} \; \ln ( 1 + \sqrt{2} ) = 0.440687
\ee
as
\be
\label{34}
\chi(g) ~ \propto ~ ( g_c - g)^{-\gm} \;   ,
\ee
with the critical index $\gamma = 7/4$. The weak-interaction or high-temperature 
expansion of the susceptibility yields \cite{Butera_17} the series in powers of $g$,
\be
\label{35}
 \chi(g) \simeq 1 + 4g + 12 g^2 \qquad (g\ra 0) \;  . 
\ee
Employing the method of Sec. 2, we find the critical point $g_c = 0.5$.  The 
first-order approximation for the critical index $\gamma_1 = 2$ differs from the exact 
one by $14.29 \%$ and the second-order approximation $\gamma_2 = 1.846$ is accurate
within  $5.49 \%$. Then the value of the index is estimated as
$$
 \gm^* = 1.923 \pm 0.077 \;  .
$$

Using high-temperature expansions for susceptibility \cite{Butera_17}, we have also 
calculated the critical indices $\gamma$ for the three-dimensional Ising model with 
different spins $S =1/2$, $3/2$, $5/2$, $7/2$, $1$, $2$, $3$, $4$, $5$, $\infty$. Since 
the critical index should not depend on the spin value, we averaged the results for 
different spins obtaining $\gamma^* = 1.2396 \pm 0.0621$, which is close to the index 
$\gamma = 1.234 \pm 0.005$ found by other sophisticated methods \cite{Kleinert_18}. 
Note that the technique of \cite{Yukalov_8} was employed for calculations for each spin 
separately.

The Dlog-Pad\'{e} method with the approximant $P_{1/1}$, given in equation (\ref{B13}),
does not have physical solutions.

\section{Effective viscosity of hard-sphere suspensions}

The problem of perfectly rigid spherical inclusions randomly embedded into an 
incompressible matrix is analogous to the problem of high-frequency effective viscosity 
of a hard-sphere suspension \cite{Batchelor_19,Wajnryb_20,Cichocki_21}. The viscosity,
considered as a function of the variable
\be
\label{36}
 \vp \equiv \frac{4\pi}{3} \; r_s^3 \rho \qquad 
\left ( \rho\equiv \frac{N}{V} \right ) \;  ,
\ee
in which $r_s$ is the sphere radius and $\rho$ is average density, exhibits the critical 
behaviour
\be
\label{37}
 \eta(\vp) ~ \propto ~ ( \vp_c - \vp)^{-\mu} \qquad ( \vp\ra \vp_c -0 ) \;  ,
\ee
where \cite{Stauffer_22}
$$
\vp_c = 0.637 \; , \qquad \mu = 1.7 \;   .
$$
The small $\varphi$-expansion reads as
\be
\label{38}
 \eta(\vp) \simeq 1 + \frac{5}{2}\; \vp + 5.0022 \vp^2 \qquad (\vp\ra 0) \;  .
\ee
Using our method, we find the critical point $\varphi_c = 0.666$. The critical index is 
$\mu_1 = 1.665$, with the error $2 \%$ and $\mu_2 = 1.788$, with the error $5 \%$. 
So the answer is
$$
 \mu^* = 1.726 \pm 0.06 \;  .
$$

The Dlog-Pad\'{e} method, with the approximant $P_{1/1}$, again does not provide
physical solutions.

\section{Conductivity in two-dimensional site percolation}

The problem of site percolation conductivity is studied within the framework of a 
minimal model for transport of classical particles through a random medium 
\cite{Nieuwenhuizen_23}. This minimal model, known as the Lorenz 2D gas, is
a particularly simple statistical hopping model allowing both for analytical consideration 
and numerical simulations \cite{Nieuwenhuizen_23,Frenkel_24}. It can be realized 
on a square lattice with a fraction of sites being excluded at random. The test 
particle, or tracer, walks randomly with Poisson-distributed waiting times between 
the moves. At every move the tracer attempts to jump on to one of the neighboring 
sites also selected at random. The move is accepted if the site is not excluded. 
Through the diffusion coefficient for the tracer one can express the macroscopic 
conductivity \cite{Nieuwenhuizen_23}. The diffusion ceases to exist at the critical 
density of the excluded sites. If $f$ stands for the concentration of conducting or 
not excluded sites in the Lorenz model, then $x = 1 - f$ is the concentration of 
excluded sites. In the vicinity of the site percolation threshold \cite{Grassberger_25}
the conductivity behaves as
\be
\label{39}
 \sgm(x) ~ \propto ~ ( x_c - x)^t \qquad ( x\ra x_c -0 ) \; ,
\ee
with
$$
 x_c = 0.4073 \; , \qquad t = 1.310 \; .
$$
Perturbation theory in powers of the variable $x = 1 - f$ gives \cite{Nieuwenhuizen_23} 
for the two-dimensional square lattice the expansion 
\be
\label{40}
 \sgm(x) \simeq 1 - \pi x + 1.28588 x^2 \qquad ( x\ra 0) \;  .
\ee

In our approach, we obtain the percolation threshold $x_c = 0.4305$ and the critical 
index $t_1 = 1.352$, with an error $3 \%$ and $t_2 = 1.423$, with an error $8.6 \%$. 
The final answer is
$$
 t^* = 1.388 \pm 0.036 \;  .
$$
 
The Dlog-Pad\'{e} method, with the approximant $P_{1/1}$, gives $t = 1.0896$,
which is worse than the above approximation, differing from the exact value by $17 \%$.

\section{Conductivity in three-dimensional site percolation}

The three-dimensional problem, similar to the two-dimensional one, treated in the previous 
section, exhibits the critical behaviour \cite{Kirkpatrick_26,Hofling_27,Bauer_28}   
as in equation (\ref{39}), with
$$
 x_c = 0.688 \; , \qquad t = 1.9 \;  .
$$
Perturbation theory gives
\be
\label{41}
 \sgm(x) \simeq 1 - 2.52 x + 1.52 x^2 \qquad ( x\ra 0) \;  .
\ee
Using our method, we get the critical point  $x_c = 0.761$. And for the critical index
we have $t_1 = 1.918$, with an error $0.9 \%$ and $t_2 = 1.855$, with an error $2 \%$. 
The answer is
$$
 t^* = 1.887 \pm 0.032 \;  .
$$

The Dlog-Pad\'{e} method, based on the approximant $P_{1/1}$, yields $t = 1.782$, with 
the error of $6 \%$, which is worse than the above value.

\section{Permeability of sinusoidal two-dimensional channel}

Let us consider the widely studied case of the two-dimensional channel bounded 
by the surfaces
$$
z = \pm b \; ( 1 + \ep\cos x ) \;  ,
$$
where $\varepsilon$ is termed {\it waviness}. The permeability possesses the critical 
behaviour \cite{Adler_29}, when (in the case of $b = 0.5$) it tends to zero as
\be
\label{42}
  K(\ep) \simeq 0.100035 (\ep_c - \ep)^t \qquad (\ep \ra \ep_c -0 ) \; ,
\ee
with
$$
 \ep_c = 1 \; , \qquad t = \frac{5}{2} \;  .
$$
An expression for permeability as a function of the waviness parameter can be derived 
by perturbation theory  in the form of an expansion in powers of the waviness 
\cite{Adler_29,Malevich_30}. Thus, the permeability, for $b = 0.5$, has the expansion
\be
\label{43}
 K(\ep) \simeq 1 - 3.14963\; \ep^2 + 4.08109\; \ep^4 \qquad (\ep\ra 0) \; .
\ee
With our method, we find $\varepsilon_c = 0.833$. For the critical index, we get  
$t_1 = 2.184$, with an error $12.6 \%$  and $t_2 = 2.559$, with an error $2.37 \%$.   
Thus we have
$$
t^* = 2.372 \pm 0.19 \;  .
$$

The Dlog-Pad\'{e} method, with the approximant $P_{1/1}$, results in $t = 1.884$, whose
error is $25 \%$, which is less accurate than the above value.

\section{Ground-state energy of harmonium atom}

An $N$-electron harmonium atom is described by the Hamiltonian
\be
\label{44}
  \hat H = \frac{1}{2} \sum_{i=1}^N \left ( -\nabla_i^2 + \om^2 r_i^2 \right ) +
\frac{1}{2} \sum_{i\neq j}^N \frac{1}{r_{ij} } \; ,
\ee
where dimensionless variables are used and
$$
 r_i \equiv | \br_i| \; , \qquad r_{ij} \equiv | \br_i - \br_j | \; .
$$
Here we consider a two-electron harmonium atom with $N = 2$. The ground-state 
energy for a rigid potential diverges \cite{Cioslowski_31} at large $\omega$ as
\be
\label{45}
 E(\om) \simeq 3\om \qquad (\om\ra\infty) \;  .
\ee
At a shallow harmonic potential, the energy can be expanded \cite{Cioslowski_31} 
in powers of $\omega$ giving
$$
  E(\om) \simeq \sum_{n=0}^k c_n \om^{(2+n)/3} \qquad (\om \ra 0 ) \; .
$$
In low orders, one has
\be
\label{46}
  E(\om) \simeq c_0 \om^{2/3} + c_1 \om + c_2 \om^{4/3} \qquad (\om \ra 0 ) \;  ,
\ee
with the coefficients
$$
c_0 = \frac{3}{2^{4/3}} = 1.19055 \; , \qquad
 c_1 = \frac{1}{2} \; ( 3 + \sqrt{3}) = 2.36603 \; , \qquad 
c_2 = \frac{7}{36}\; 2^{-2/3} = 0.122492 \;  .
$$
Introducing the new variable
\be
\label{47}
 x \equiv \om^{1/3} \;  ,
\ee
equation (\ref{46}) reduces to 
\be
\label{48}
 E(x^3) \simeq c_0 x^2 ( 1 + a_1 x + a_2 x^2 ) \qquad ( x \ra 0 ) \;  ,
\ee
with the coefficients
$$
a_1 = \frac{c_1}{c_0} = 1.98734 \; , \qquad  a_2 = \frac{c_2}{c_0} = 0.102887 \; .
$$
Employing our method, we find the large $\omega$ behaviour
\be
\label{49}
E_1^*(\om) \simeq 2.322\; \om^{\bt_1} \qquad (\om \ra \infty)
\ee
and
\be
\label{50}
 E_2^*(\om) \simeq 1.906\; \om^{\bt_2} \qquad (\om \ra \infty) \; ,
\ee
where
$$
 \bt_1 = 1.018 \; , \qquad \bt_2 =  1.079 \;  .
$$
The error of $\beta_1$ is $1.8 \%$ and of $\beta_2$, it is $7.9 \%$. The accuracy 
is compared with the known numerical data \cite{Matito_32}. The resulting effective 
critical index at infinity is
$$
 \bt^* = 1.049 \pm 0.031 \;  .
$$

The Dlog-Pad\'{e} method, with the approximant $P_{1/1}$, has no physical solutions.

\section{Compressibility factor of hard-sphere fluids}

The state of hard-sphere fluids is described by the compressibility factor
\be
\label{51}
 Z = \frac{P}{\rho k_B T} = Z(y) \qquad 
\left ( y \equiv \frac{\pi\rho}{6}\; a_s^3 \right ) \;  ,
\ee
in which $P$ is pressure, $\rho$ is density, $T$ is temperature, $a_s$ is the sphere 
diameter, and $y$ is called packing fraction \cite{Mulero_35}. 

The compressibility factor exhibits critical behaviour at a finite critical point. This 
behavior has been found from phenomenological equations 
\cite{Wu_36,Tian_37,Tian_38,Wang_48} as
\be
\label{52}
Z(y) \simeq 2 (y_c - y)^{-t} \qquad ( y \ra y_c -0) \; ,
\ee
with the fitted parameters $y_c = 1$ and $t = 3$, although these are not asymptotically 
exact values.  

For low packing fraction, the compressibility factor is represented by the virial expansion
\be
\label{53}
 Z(y) \simeq 1 + 4y + 10y^2 \;  .
\ee
Using the method of Sec. 2, we find, with $y_c = 1$, the indices $t_1 = 4$ and $t_2 = 4.686$. 
Therefore our prediction for the critical index is
$$
 t^* = 4.343 \pm 0.34 \;  .
$$

The Dlog-Pad\'{e} method, with the approximant $P_{1/1}$, gives $t = 4$.

\section{Expansion factor of polymer chain}

The expansion factor of a polymer chain, as a function of the dimensionless coupling
parameter $g$, can be expressed by the phenomenological equation 
\cite{Muthukumar_39,Muthukumar_40}
\be
\label{60}
 \al(g) = ( 1 + 7.52 g + 11.06 g^2 )^{0.1772} \;  .
\ee
At large $g$, this gives
\be
\label{61}
 \al(g) \simeq 1.531 g^\bt \qquad ( g \ra \infty ) \;  ,
\ee
with the critical index at infinity $\beta = 0.3544$. One also considers the critical index
\be
\label{62}
 \nu \equiv \frac{1}{2} \left ( 1 + \frac{\bt}{2} \right )  
\ee
that here is $\nu = 0.5886$. Other numerical calculations \cite{Li_41} give $\nu = 0.5877$.
At small $g$, perturbation theory yields \cite{Muthukumar_39,Muthukumar_40} the expansion
\be
\label{63}
\al(g) \simeq 1 + \frac{4}{3} \; g - 2.075385 g^2  \qquad ( g \ra 0 ) \; .
\ee

By the method of Sec. 2, we obtain the critical behaviour
\be
\label{64}
 \al_1^*(g) \simeq 1.544 g^{\bt_1} \qquad ( g \ra \infty ) \;  ,
\ee
with the critical indices 
$$
\bt_1 = 0.2999 \; , \qquad \nu_1=0.5750 \;   ,
$$
the error being just $0.023 \%$. And the error of $\nu_2 = 0.5878$ is only $0.0013 \%$. 
In this way,
$$
\nu^* = 0.5814 \pm 0.006 \; .
$$

The Dlog-Pad\'{e} method, with the approximant $P_{1/1}$, does not possess physical 
solutions.

\section{Sedimentation coefficient of rigid spheres}

Sedimentation is a fundamental problem of studying how a suspension moves under 
gravity. The considered dispersion is build of small rigid spheres with random positions 
falling through Newtonian fluid under gravity. The mixture of solid particles and the 
fluid in a container is assumed to be homogeneous. The particles settle out under gravity 
at a rate depending, in particular, on concentration originating from hydrodynamic 
interactions between particles. The basic quantity of interest is the sedimentation velocity 
$U$, which is the averaged velocity of suspended particles, measured with respect to
the velocity $U_0$ with which a single particle would move in the suspending fluid under 
the given force field in the absence of any other particles. This ratio is termed the 
collective mobility or sedimentation coefficient. The dependence of the collective mobility 
at low packing fractions is similar to the single-particle mobility, but quickly decreases 
at high packing fractions. The problem of sedimentation reminds that of Darcy flow in a 
porous medium, although their relation is not simple, because the physics of particle 
interactions is rather different. More details on the physics of sedimentation can be 
found in the paper by Batchelor \cite{Batchelor_42}.

The dimensionless sedimentation velocity $u \equiv U/U_0$ is considered as a function
of the packing fraction $f$. This velocity exhibits the critical behaviour 
\be
\label{65}
u(f) ~ \propto ~ (1 - f)^\bt \qquad ( f \ra 1 - 0 )
\ee
at the critical point $f_c = 1$. The critical index, however, has been defined 
differently by different authors. Thus Batchelor \cite{Batchelor_42} gives $\beta = 5$. 
While other authors \cite{Beenakker_43,Brady_44,Ladd_45,Hayakawa_46} suggest $\beta = 3$. 
Below we find the critical index being based on the expansion derived by Cichocki et al 
\cite{Cichocki_47},
\be
\label{66}
 u(f) \simeq 1 - 6.546 f + 21.918 f^2 \qquad ( f \ra 0 ) \; .
\ee
 
By the method of Sec. 2, setting $f_c = 1$, we find $\beta_1 = 3.0438$ and 
$\beta_2 = 3.5660$. Therefore we predict the critical index
$$
 \bt^* = 3.3049 \pm 0.26 \;  .
$$

The Dlog-Pad\'{e} method, with the approximant $P_{1/1}$, has no physical solutions.

\section{Ground-state energy of Schwinger model}

The Schwinger model \cite{Schwinger_48,Banks_49} represents Euclidean quantum 
electrodynamics with a Dirac fermion field, which is formulated as a lattice gauge theory
in $(1+1)$ dimensions. It enjoys many properties in common with quantum chromodynamics, 
such as confinement, chiral symmetry breaking, and charge shielding, because of which 
it has become a standard test bed for the study of numerical techniques. Here we consider 
the model corresponding to a vector boson of mass $M(x)$ as a function of the 
dimensionless variable $x \equiv m/g$, with $m$ being the electron mass and $g$ being 
the coupling parameter. The ground-state energy is given by the expression $E - 2m$. 

The energy, as a function of $x$, increases with $x$, reaching the asymptotic value
\cite{Striganesh_50,Coleman_51,Hamer_52,Hamer_53}
\be
\label{67}
 E(x) \simeq 0.6418\; x^\bt \qquad ( x \ra \infty) \;  ,
\ee
with the critical index at infinity $\beta = 1/3$. 

At small $x$, there is the expansion \cite{Striganesh_50,Carrol_54,Vary_55,Adam_56}
\be
\label{68}
 E(x) \simeq 0.5642 ( 1 - 0.38816\; x + 0.338001\; x^2 ) \qquad ( x \ra 0) \;  .
\ee

Using the method of Sec. 2, we find $\beta_1 = - 0.2868$ that differs from the exact
$\beta = - 1/3$ by $13.96 \%$ and $\beta_2 = - 0.3360$ differing from the exact value
by $0.8 \%$. Thus, we get
$$
\bt^* = 0.311 \pm 0.02 \; .
$$

The Dlog-Pad\'{e} method, with the approximant $P_{1/1}$, does not have physical solutions.

\section{Convergence of approximants for critical indices}

In the present section, we illustrate the numerical convergence of root approximants
applied for calculating critical indices. As an example, we consider the pressure $P(x)$ of 
a fluctuating fluid membrane \cite{Seifert_58} as a function of stiffness $x$. This example 
is of special interest, since it cannot be treated by Pad{\'e} approximants \cite{Gluzman_12}. 

The pressure can be represented in the form
\be
\label{69}
  P(x) = \frac{\pi^2}{8x^2} \; f(x) \; .
\ee
It has been found by Monte Carlo simulations \cite{Gompper_59} that the function $f(x)$ 
diverges at infinity as
\be
\label{70}
 f(x) \simeq 0.06468 x^2 \qquad ( x \ra \infty) \;  .
\ee
Hence, this function exhibits the critical behaviour at infinity with the critical index 
$\beta = 2$.

At weak stiffness, the function $f(x)$ can be found \cite{Kastening_60} by perturbation 
theory with respect to the stiffness, yielding the expansion
\be
\label{71}
 f(x) \simeq \sum_{n=0}^k a_n x^n \qquad ( x \ra 0 ) \;  ,
\ee
with the coefficients
$$
 a_0 = 1 \; , \qquad a_1 = \frac{1}{4} \; , \qquad a_2 =\frac{1}{32} \; , \qquad 
a_3 = 2.176347\times 10^{-3} \;  ,
$$
$$
a_4 = 0.552721\times 10^{-4} \; , \qquad a_5 = - 0.721482 \times 10^{-5} \; , \qquad
a_6 = - 1.777848 \times 10^{-6} \; ,
$$
which can be complemented by two more coefficients $a_7 = a_8 = 0$. 

We construct the root approximants
\be
\label{72}
f_k^*(x) = \left ( \left ( \left (  ( 1 + A_1 x)^2 + A_2 x^2 \right )^{3/2}
+ A_3 x^3 \right )^{4/3} + \ldots + A_k x^k \right )^{\bt_k/k} \;   ,
\ee
defining the parameters $A_j$ from the accuracy-through-order procedure, as is 
explained in Sec. 1. This gives the large-stiffness asymptotic forms
\be
\label{73}
  f_k^*(x) \simeq B_k x^\bt_k \qquad ( x \ra \infty ) \; ,
\ee
where the amplitudes $B_k = B_k(\beta_k)$ are 
\be
\label{74}
 B_k =  \left ( \left (  ( A_1^2 + A_2)^{3/2} + A_3 \right )^{4/3}
+ \ldots + A_k \right )^{\bt_k/k} \;  .
\ee
In order to define the critical index $\beta_k$, we follow Sec. 2 and analyze the
differences
\be
\label{75}
 \Dlt_{kn}(\bt_k) = B_k(\bt_k) - B_n(\bt_k) \;  .
\ee
Composing the sequences $\Delta_{kn} = 0$, we find the related approximate values 
$\beta_k$ for the critical indices. It is possible to investigate different sequences
of the conditions $\Delta_{kn} = 0$, the most logical from which are the sequences
of $\Delta_{k,k+1} = 0$ and of $\Delta_{k8} = 0$, with $k = 1,2,3,4,5,6,7$. The
results, presented in Table 1, show good numerical convergence of the approximate
critical indices $\beta_k$ to the exact value $\beta = 2$.  

The Dlog-Pad\'{e} method, with the approximant $P_{1/1}$, again does not provide 
physically acceptable solutions. And if we employ the approximant $P_{N/N+1}$, this
results in the sequence of the critical indices $-0.08095$, $2.5188$, $-5.3603$, and 
$-2.19958$ for $N = 1,2,3,4$, respectively. As is evident, such a sequence is neither
convergent nor reasonable.

\begin{table}
\centering
\begin{tabular}{|c|c|c|} \hline
$\bt_k$  & $\Dlt_{k,k+1}(\bt_k)=0$  &  $\Dlt_{k8}(\bt_k)=0$   \\ \hline
$\bt_1$  &       24.9036            &   2.5052                \\ \hline
$\bt_2$  &        4.9344            &   2.4701                \\ \hline
$\bt_3$  &        3.4791            &   2.3887                \\ \hline
$\bt_4$  &        2.8459            &   2.2970                \\ \hline
$\bt_5$  &        2.3983            &   2.2018                \\ \hline
$\bt_6$  &        2.2040            &   2.1289                \\ \hline
$\bt_7$  &        2.0645            &   2.0645              \\ \hline  
\end{tabular}
\caption{Critical indices $\bt_k$ for the problem of Sec. 14, obtained from the optimization
conditions $\Dlt_{kn}(\bt_k)=0$. The sequences of $\bt_k$ demonstrate numerical convergence 
to the exact value $\bt=2$.}
\label{tab}
\end{table}

\section{Concluding comments}

The method of self-similar root approximants, proved earlier to provide accurate
{\it interpolation} for the sought function, when both the asymptotic expansions at 
small and large variables are known. Now the method is generalized to the case when 
only a small-variable expansion is available and the function can display critical 
behaviour at a finite critical point or at infinity. We show how, having in hands only 
a small-variable expansion, one can construct {\it extrapolation} formulas for the 
sought function and to find its critical index. The method of defining critical indices 
is illustrated by a large set of examples for various physical problems. It is shown 
that the suggested approach makes it straightforward to calculate critical indices, 
with a good accuracy, even when just a few terms of small-variable expansions are 
available. When a number of terms in the small-variable expansion is given, the method 
demonstrates numerical convergence to the exact indices, if the latter are known. 

Following the idea of the approach, it is admissible to realize different variants of the
calculational scheme. For instance, instead of treating the sought function $f(x)$,
and its related $k$-order expansion $f_k(x)$, one can consider the inverse 
expression $h_k(x) \equiv 1/f_k(x)$. Then, constructing the root approximants $h_k^*(x)$,
it is easy to return back to $f_k^*(x) = 1/h_k^*(x)$. 

The other possibility can be useful, when the critical behaviour occurs at a finite critical 
point $x_c$. By the change of the variable 
$$
 z = \frac{x}{x_c-x} \; , \qquad x = \frac{x_c z}{1+z} \;   
$$
one shifts the critical point to infinity. And then finds the critical index as is explained 
in Sec. 2. 
  
We have checked both these variants for the examples considered and found that 
they give close results for the critical indices, as compared to the direct method
exposed above. 

In conclusion, the developed method of defining critical indices is general and can be 
applied to different physical problems. The method works well even when other methods,
such as that of Pad\'{e} approximants, are not applicable. 

\vskip 5mm

{\bf Acknowledgment}

\vskip 2mm
One of the authors (V.I.Y.) is grateful for discussions to E.P. Yukalova.

\end{document}